\PassOptionsToPackage{unicode}{hyperref}
\PassOptionsToPackage{hyphens}{url}
\PassOptionsToPackage{dvipsnames,svgnames,x11names}{xcolor}
\documentclass[12pt]{article}

\usepackage{latexsym,amssymb,amsmath,amsbsy,amsopn,amstext,amsthm,bm,amsxtra,color,multicol,calc,ifpdf}
\usepackage{graphicx}
\usepackage{diagbox}  
\usepackage{enumerate}
\usepackage{fancyhdr}
\usepackage{float}
\usepackage{listings}
\usepackage{multirow}
\usepackage{makeidx}
\usepackage{multirow}
\usepackage{booktabs}
\usepackage{ulem}
\usepackage{mathrsfs}
\usepackage{algorithm}
\usepackage{algorithmic}

\usepackage{iftex}
\ifPDFTeX
  \usepackage[T1]{fontenc}
  \usepackage[utf8]{inputenc}
  \usepackage{textcomp} 
\else 
  \usepackage{unicode-math}
  \defaultfontfeatures{Scale=MatchLowercase}
  \defaultfontfeatures[\rmfamily]{Ligatures=TeX,Scale=1}
\fi
\usepackage{lmodern}
\ifPDFTeX\else  
\fi
\IfFileExists{upquote.sty}{\usepackage{upquote}}{}
\IfFileExists{microtype.sty}{
  \usepackage[]{microtype}
  \UseMicrotypeSet[protrusion]{basicmath} 
}{}
\makeatletter
\@ifundefined{KOMAClassName}{
  \IfFileExists{parskip.sty}{%
    \usepackage{parskip}
  }{
    \setlength{\parindent}{0pt}
    \setlength{\parskip}{6pt plus 2pt minus 1pt}}
}{
  \KOMAoptions{parskip=half}}
\makeatother
\usepackage{xcolor}
\makeatletter
\ifx\paragraph\undefined\else
  \let\oldparagraph\paragraph
  \renewcommand{\paragraph}{
    \@ifstar
      \xxxParagraphStar
      \xxxParagraphNoStar
  }
  \newcommand{\xxxParagraphStar}[1]{\oldparagraph*{#1}\mbox{}}
  \newcommand{\xxxParagraphNoStar}[1]{\oldparagraph{#1}\mbox{}}
\fi
\ifx\subparagraph\undefined\else
  \let\oldsubparagraph\subparagraph
  \renewcommand{\subparagraph}{
    \@ifstar
      \xxxSubParagraphStar
      \xxxSubParagraphNoStar
  }
  \newcommand{\xxxSubParagraphStar}[1]{\oldsubparagraph*{#1}\mbox{}}
  \newcommand{\xxxSubParagraphNoStar}[1]{\oldsubparagraph{#1}\mbox{}}
\fi
\makeatother

\usepackage{longtable,booktabs,array}
\usepackage{calc} 
\usepackage{etoolbox}
\makeatletter
\patchcmd\longtable{\par}{\if@noskipsec\mbox{}\fi\par}{}{}
\makeatother
\IfFileExists{footnotehyper.sty}{\usepackage{footnotehyper}}{\usepackage{footnote}}
\makesavenoteenv{longtable}
\usepackage{graphicx}
\makeatletter
\def\maxwidth{\ifdim\Gin@nat@width>\linewidth\linewidth\else\Gin@nat@width\fi}
\def\maxheight{\ifdim\Gin@nat@height>\textheight\textheight\else\Gin@nat@height\fi}
\makeatother
\setkeys{Gin}{width=\maxwidth,height=\maxheight,keepaspectratio}
\makeatletter
\def\fps@figure{htbp}
\makeatother

\makeatletter
\@ifpackageloaded{caption}{}{\usepackage{caption}}
\AtBeginDocument{%
\ifdefined\contentsname
  \renewcommand*\contentsname{Table of contents}
\else
  \newcommand\contentsname{Table of contents}
\fi
\ifdefined\listfigurename
  \renewcommand*\listfigurename{List of Figures}
\else
  \newcommand\listfigurename{List of Figures}
\fi
\ifdefined\listtablename
  \renewcommand*\listtablename{List of Tables}
\else
  \newcommand\listtablename{List of Tables}
\fi
\ifdefined\figurename
  \renewcommand*\figurename{Figure}
\else
  \newcommand\figurename{Figure}
\fi
\ifdefined\tablename
  \renewcommand*\tablename{Table}
\else
  \newcommand\tablename{Table}
\fi
}
\@ifpackageloaded{float}{}{\usepackage{float}}
\floatstyle{ruled}
\@ifundefined{c@chapter}{\newfloat{codelisting}{h}{lop}}{\newfloat{codelisting}{h}{lop}[chapter]}
\floatname{codelisting}{Listing}

\makeatother
\makeatletter
\@ifpackageloaded{caption}{}{\usepackage{caption}}
\@ifpackageloaded{subcaption}{}{\usepackage{subcaption}}
\makeatother

\ifLuaTeX
  \usepackage{selnolig}  
\fi
\usepackage[]{natbib}
\usepackage{bookmark}

\IfFileExists{xurl.sty}{\usepackage{xurl}}{} 
\hypersetup{
  pdftitle={Title},
  pdfauthor={Author 1; Author 2},
  pdfkeywords={3 to 6 keywords, that do not appear in the title},
  colorlinks=true,
  linkcolor={blue},
  filecolor={Maroon},
  citecolor={Blue},
  urlcolor={Blue},
  pdfcreator={LaTeX via pandoc}}



\newcommand{\abs}[1]{\left\lvert#1\right\rvert}
\newcommand{\norm}[1]{\left\lVert#1\right\rVert}

\newtheorem{thm}{Theorem}
\newtheorem{prop}{Proposition}%

\newtheorem{assum}{Assumption}
\newtheorem{exam}{Example}%
\newtheorem{Rmk}{Remark}%

\usepackage[doublespacing]{setspace}
\usepackage[fontsize=12pt]{fontsize}

\newcommand{\anon}{1}

\begin{document}

\def\spacingset#1{\renewcommand{\baselinestretch}%
{#1}\small\normalsize} \spacingset{1}


\if1\anon
{
  \title{\bf Structural Change Detection in Dynamic Systems}
  \author{Wei Zhang\hspace{.2cm}\\
    School of Management, Department of Statistics and Data Science,\\
    Fudan University, Shanghai, China\\
    and \\
    Fang Yao\thanks{Fang Yao is the corresponding author. \href{Email:fyao@math.pku.edu.cn}{fyao@math.pku.edu.cn}} \\
    School of Mathematical Sciences, Center for Statistical Science,\\ 
    Peking University, Beijing, China}
  \maketitle
} \fi

\if0\anon
{
  \bigskip
  \bigskip
  \bigskip
  \begin{center}
    {\LARGE\bf Structural Change Detection in \\ Dynamic Systems}
\end{center}
  \medskip
} \fi

\bigskip
\begin{abstract}
Structural changes often arise in real-world dynamic systems due to external interventions or environmental shifts, such as policy changes in epidemiology or climate forcing in environmental science. In this paper, we propose a unified framework for detecting and localizing structural changes in dynamic systems governed by ordinary differential equations. Unlike existing methods that assume  mean or linear trend changes, our approach accommodates complex, nonlinear dynamics with both stable and diverging trajectories. We develop a new test statistic that combines residual-based discrepancy and normalized parameter contrast, capturing evidence for structural changes from both model fit and parameter shifts. Candidate structural changes are efficiently screened using a multiscale seeded-narrowest-over-threshold algorithm with a data-driven thresholding strategy. To refine selections and control false discoveries, we introduce a false discovery rate control procedure that leverages order-preserved sample splitting and symmetric contrast calibration. Theoretical guarantees are established, including detection consistency, near-minimax localization accuracy, and valid FDR control under weak dependence. Extensive simulations demonstrate superior performance over existing methods in both accuracy and FDR control. Applications to real-world data sets, including COVID-19 dynamics and global temperature trends, highlight the practical relevance and broad applicability of our method.
\end{abstract}

\noindent%
\vfill

\newpage
\spacingset{1.8} 

\setlength{\abovedisplayskip}{2.8pt}
\setlength{\belowdisplayskip}{2.8pt}
\setlength{\abovedisplayshortskip}{2.8pt}
\setlength{\belowdisplayshortskip}{2.8pt}

\section{Introduction}
\label{sec:intro}

Dynamic systems play a central role in modeling complex temporal behaviors across a wide range of scientific and practical domains, such as climate modeling, spread of disease and etc \citep{Miao09, Michael20}. In reality, these systems often experience structural changes over time, significantly altering their behavior and complicating predictive modeling. For example, an epidemiological model may experience structural changes due to the implementation or lifting of public health interventions like lockdowns or vaccination campaigns \citep{Brauner21}. Similarly, the global temperature model can undergo structural changes influenced by varying factors such as solar activity and atmospheric changes \citep{Benestad09}. Reliably detecting these structural changes is critical for enhanced modeling accuracy, deeper comprehension of system behavior, and informed decision-making in practical applications.

The general form of these dynamic systems is often mathematically expressed as a set of ordinary differential equations (ODEs):
\begin{equation}
\label{eq:ode}
\Dot{u}(t) = \frac{du(t)}{dt} = f(u, t; \theta), \quad  u(0) = \gamma,
\end{equation}
where the state vector $u(t) \in \mathbb{R}^d$ represents system outputs evolving continuously over time $t$ and $\theta \in \mathbb{R}^q$ contains any parameters defining the system whose values are not known from experimental data, theoretical considerations or other sources of information. The known function $f(\cdot)$ explicitly describes the relationship between the system outputs and their time derivatives. If either there are no observations at $t=0$, or the observations contain measurement errors, then the initial value $\gamma$ is also unknown.

As previously mentioned, dynamic systems in real-world applications are frequently subject to structural changes driven by various external influences. In most cases, these structural changes are reflected in the unknown parameter $\beta = (\theta^\top, \gamma^\top)^{\top}$ but do not change the form of $f(\cdot)$. This is because $f(\cdot)$ typically encodes the underlying physical, biological, or theoretical principles governing the system, which usually remain invariant over time. The parameters $\beta$, on the other hand, capture system-specific conditions such as intensities, rates, sensitivities, or initial states, which are more likely to be affected by external interventions or environmental changes \citep{Keeling08, Strogatz15}. For instance, in an epidemiological model, the infection rate may vary due to policy changes or behavioral shifts, altering parameter values, while the basic structure of disease transmission remains unchanged. Therefore, modeling structural changes as parameter shifts preserves the interpretability and scientific validity of the system while accommodating the non-stationary nature of real-world data.

Hereafter, we focus on the following observation model: for $k = 0, \dots, K_n$ and $i = 1, \dots, n$,
\begin{equation}
\label{eq:ode_change}
y(t_i) = u(t_i; \beta^*_k) + \varepsilon_i, \quad \tau_k^* < i \leq \tau_{k+1}^*,
\end{equation}
where $K_n$ is the total number of structural changes and $\tau_k^*$ denotes the index of the $k$-th change location. Given the continuous nature of the underlying process, it indicates that parameters transition from $\beta_{k-1}^*$ to $\beta_{k}^*$ in $[t_{\tau_k^*}, t_{\tau_{k}^*+1}]$, $k = 1, \dots, K_n$. The noise terms $\varepsilon_i = (\varepsilon_{i1}, \dots, \varepsilon_{id})^{\top}$ are assumed to be i.i.d. and sub-exponentially distributed. For notational convenience, we define $\tau_0^* = 0$ and $\tau_{K_n + 1}^* = n$, with the condition $\beta_{k-1}^* \neq \beta_{k}^*$ to ensure genuine structural shifts. We remark that when intermediate observations are unavailable, the exact timing of such changes is inherently unidentifiable. Thus, our focus is on detecting the discrete structural change locations $\tau_k^*$, $k = 1, \dots, K_n$ rather than pinpointing the exact transition times.
\textcolor{black}{This problem is closely related to multiple change-point (MCP) detection that has been studied extensively for mean change-point models with methods including wild binary segmentation and related procedures \citep{Fryzlewicz14, Kovacs22}. Existing work beyond the mean model includes changes in linear trends \citep{Bai98, Fearnhead19, Tibshirani}, generalized piecewise linear or polynomial signals \citep{Baranowski19}, linear regression settings \citep{Fryzlewicz02042024}, parametric MCP problems reducible through score-type transformations \citep{Chen03042023}, and nonparametric jump regression \citep{Wang95, Qiu15}.} 

\textcolor{black}{Compared with these frameworks, our setting presents distinct statistical and computational challenges, where the signal is not directly observed but arises as the solution of an ODE with unknown and potentially time-varying parameters. As a result, structural changes need not appear as discontinuities or cusps in the trajectory itself, but may instead be expressed through changes in its higher-order derivative structure, while inference must additionally account for parameter estimation and numerical approximation within the dynamical system. Related formulation for indirect noisy observations in stochastic differential equations \citep{Zeevi06, Markus2025AOS} do not cover this setting as the analysis is restricted to jump discontinuities or finite observation horizons. Locally stationary processes \citep{Dette15} provide another framework for modeling time-varying stochastic properties, but our underlying dynamical system is usually nonstationarity rather than slowly varying characteristics. These differences place our problem outside the scope of existing MCP methodologies and motivate the development of new tools for structural change detection under implicit dynamical models.}

In this paper, we propose a new framework for detecting and localizing structural changes in dynamic systems governed by ODEs. The method is built around a test statistic that combines two sources of evidence: a residual-based discrepancy and a normalized contrast between parameter estimates on adjacent segments, ensuring both robustness and statistical power under complex model structures. Candidate structural changes are efficiently identified via the seeded-narrowest-over-threshold (SNOT) algorithm, which performs a multiscale scan over a deterministic set of seeded intervals and selects structural changes that exceed a data-driven threshold derived from the distribution of these intervals. To refine the candidate set, we develop a False Discovery Rate (FDR) control procedure based on order-preserving sample splitting and symmetric contrast calibration. To accommodate systems with diverging trajectories, such as those found in linear trend models, we further introduce a parameter-specific adaptive scaling strategy that restores the asymptotic normality of contrast statistics across varying dynamic regimes. This unified framework delivers consistent detection and valid post-selection inference while remaining broadly applicable to a wide class of complex ODE systems.

The main contributions of this work are as follows. First, we develop a unified framework for detecting and inferring structural changes in dynamic systems, capable of handling both stable and diverging trajectories without requiring closed-form solutions of the governing ODEs. Second, we introduce a novel data-driven procedure to determine the threshold for initial screening in the SNOT algorithm, enabling efficient and adaptive candidate selection. Third, we propose a principled FDR control method that ensures valid inference, offering finite-sample guarantees and consistent selection even under weak dependence and asymmetric test statistics. Fourth, we establish comprehensive theoretical results, including detection consistency, near-minimax localization error bounds, and asymptotically valid FDR control. We remark here that the FDR control properties are built for the whole detection procedure, while most existing theoretical analyses of post-detection procedures are based on assumptions of detection results \citep{Chen03042023}. Finally, we validate our method through extensive simulation studies, demonstrating superior performance over existing approaches in terms of power, false discovery control, and localization precision. The practical utility of our approach is further illustrated through two real data applications: modeling the spread of COVID-19 and analyzing global temperature dynamics.

The rest of the paper is organized as follows. Section \ref{sec:method} introduces the proposed framework for detecting and localizing structural changes in dynamic systems, as well as the modifications for diverging trajectories. In Section \ref{sec:thm}, we derive the detection consistency of the SNOT algorithm, and prove that the proposed FDR control procedure can adequately control the false discoveries without missing true structural changes. Comprehensive simulation studies on bounded and diverging systems are conducted in Section \ref{sec:simulation}, and two real data examples for structural change detection are presented in Section \ref{sec:application}. Concluding remarks are provided in Section \ref{sec:discussion}. More discussion on the definition of FDR, theoretical details, along with additional simulation settings and results are provided in the Supplementary Material for space economy.

\section{Multiple Structural Change Detection in ODEs}
\label{sec:method}
In this section, we formally describe the whole detection procedure. We first generate a sequence of generic intervals by seeded binary segmentation \citep{Kovacs22}, and for each interval, we determine the existence of changes and select a candidate structural change with our proposed test statistics, followed by utilizing a searching algorithm and FDR control procedure to derive the final estimation of structural changes. The detailed test statistic is introduced in the subsection \ref{subsec:statistics}. The proposed searching algorithm and FDR control procedure are provided in subsections \ref{subsec:algorithm} and \ref{subsec:FDR}. In addition, the modifications for diverging trajectories are provided in subsection \ref{subsec:growth}.

\subsection{Test statistic}
\label{subsec:statistics}

Suppose we observe a sequence of measurements \( \{y(t_i)\}_{i=1}^n \) generated from the dynamic model \eqref{eq:ode_change}, which allows for potential changes in the underlying system parameters \( \beta \). Our objective is to develop a procedure that determines whether a change occurs within a given generic interval \( (s, e]\), where \( s, e \in \{1, \dots, n\} \), and if so, to estimate the corresponding location. To this end, we construct a test statistic based on two key components: (i) a normalized measure of parameter contrast across subintervals which assesses whether a meaningful change has occurred, and (ii) a residual-based discrepancy statistic which identifies the most plausible location of a structural break when such a change is present.

We begin by considering a integer-valued candidate split point \( r \in (s, e) \) that partitions the interval into two subsegments: \( (s, r] \) and \( (r, e] \). Distinct from the mean change or linear regression setting, the changes can be  directly expressed through the model structure. By contrast, the parameters governing the ODEs are implicit as the solution $u$ are unobserved and contaminated with noise. A typical remedy to recover $u$ is smoothing-based procedures \citep{Cao07, Tan24, Markus2025AOS} that incorporate the ODE/SDE through a penalty term or, when available, a mild-solution representation. However, these type of methods can only deal with problems with densely observed data filling within a finite domain. In reality, many dynamic systems evolve along time that extends with more observations. Although one may rescale the increasing domain to a finite interval, this might alter the underlying dynamics and dependence among observations. Therefore, to keep fidelity to the governing ODEs, we take advantage of existing ODE solvers, combining powerful numerical tools with change detection problems. To this end, for each subsegment, we estimate the system parameters by solving the following least squares problems:
\[
\hat{\beta}_{(s, r]} = \arg\min_{\beta} \sum_{i=s+1}^{r} \left\| y(t_i) - \hat{u}(t_i; \beta) \right\|_2^2, \quad
\hat{\beta}_{(r, e]} = \arg\min_{\beta} \sum_{i=r+1}^{e} \left\| y(t_i) - \hat{u}(t_i; \beta) \right\|_2^2,
\]
where \( \hat{u}(t; \beta) \) denotes the numerical solution of the ODE system at time \( t \). 

\begin{Rmk}
   \textcolor{black}{Unlike smoothing-based reconstruction from noisy data, the approximation $\hat{u}$ depends only on the ODE and therefore introduces only numerical error, which can be controlled by standard solver tolerances. This is useful for analyzing the properties of structural change detection, where careful control of stochastic fluctuations is important for the tail behavior of the test statistics. By avoiding data-driven smoothing, one preserves the separation of the odd and even subsamples in the subsequent FDR procedure, and can extend this strategy easily to more general nonlinear regression models.}
\end{Rmk}
To quantify the extent of change across the candidate split point, we construct a normalized parameter contrast as
\[
\Delta_{(s, e]}(r) = \sqrt{ \frac{(e - r)(r - s)}{e - s} } \left\| \hat{\beta}_{(s, r]} - \hat{\beta}_{(r, e]} \right\|_2.
\]
The normalization factor adjusts for the relative lengths of the subintervals, ensuring comparability across different values of \( r \). We then define the maximum contrast over the interval as $\Delta_{(s, e]} = \max_{s + p + 1 \leq r \leq e - p - 1} \Delta_{(s, e]}(r)$, where \( p=q+d \) is the dimension of the parameter \( \beta \), and the exclusion of $p+1$ boundary points ensures that both local estimators can be estimated from nearby data. A large value of \( \Delta_{(s, e]} \) suggests that the interval $(s, e]$ includes a substantial change in the system parameters. Thus, if the parameter contrast \( \Delta_{(s, e]} \) does not exceed a predetermined threshold \( \zeta_n > 0 \), we conclude that the interval \( (s, e] \) does not contain a structural change and refrain from further evaluation. Otherwise, if \( \Delta_{(s, e]} > \zeta_n \), we proceed to locate the structural change.

To identify the most plausible change location within \( (s, e] \), we employ a residual-based loss function:
\[
R_{(s, e]}(r) = \sum_{i=s+1}^{r} \left\| y(t_i) - \hat{u}(t_i; \hat{\beta}_{(s, r]}) \right\|_2^2 + \sum_{i=r+1}^{e} \left\| y(t_i) - \hat{u}(t_i; \hat{\beta}_{(r, e]}) \right\|_2^2.
\]
This statistic captures the total residual sum of squares when separate models are fit on either side of the candidate split. In contrast to classical CUSUM-type approaches, which measure cumulative deviations, this statistic measures the reduction in residual error due to introducing a structural change at $r$. The estimated location of structural change within the interval is then selected as $\hat{r}_{(s, e]} = \arg\min_{s + p + 1 \leq r \leq e - p - 1} R_{(s, e]}(r)$, which is the location that improves the model fit to the most extent.

To conclude, we can define the test statistic as
\[
T_{(s, e]}(r) = R_{(s, e]}(r)\cdot\mathbf{1}\left\{\Delta_{(s, e]} > \zeta_n\right\} + \infty \cdot \mathbf{1}\left\{\Delta_{(s, e]} \leq \zeta_n\right\}
\]
The threshold \( \zeta_n \) serves to exclude intervals with insufficient evidence of change by setting the test statistics to be infinite. Then a small $T_{(s, e]}(r)$ implies that $(s,e]$ contains a structural change near $r$. This design provides statistical rigor by avoiding over-detection in homogeneous regions while retaining accuracy in detecting structural change locations. In the following subsection, we build on this statistic to develop a searching algorithm for multiple structural change detection.

\begin{Rmk}
A related test statistic is the generalized likelihood ratio (GLR) statistic discussed in \cite{Baranowski19}. Conceptually, the GLR constructs a likelihood ratio test at each candidate location, while our proposed statistic is based on a Wald-type test. Under standard regularity conditions, these two approaches are asymptotically equivalent. In particular, when the noise is Gaussian, both statistics tend to identify the same most plausible change locations. However, the GLR statistic requires an additional fit over the full interval $(s, e]$, which incurs a non-negligible computational cost during the scanning procedure. In contrast, our test statistic is more computationally efficient and also facilitates the theoretical analysis. Empirical comparisons presented in Figure S1 of the Supplementary Material show that the GLR statistic appears slightly more conservative than the proposed one. This may be attributed to the fact that, in the simulation settings, the changes in the trajectory are less pronounced than the underlying parameter shifts.

\end{Rmk}

\subsection{Searching with screening}
\label{subsec:algorithm} 

\textcolor{black}{Based on the candidate structural change $\hat{r}_{(s,e]}$ and its corresponding test statistic $T_{(s,e]}(\hat{r}_{(s,e]})$ computed for each seeded interval $(s,e]$, we construct a set of candidate change points using a seeded-narrowest-over-threshold (SNOT) search procedure. The method combines a deterministic seeded interval design with the narrowest-over-threshold principle to identify potential structural changes. Specifically, after discarding intervals with no detectable change, the algorithm iteratively selects the shortest interval whose test statistic exceeds a threshold. When multiple intervals share the same length, the interval with the largest parameter contrast $\Delta_{(s,e]}$ is selected. Once a change point is accepted, overlapping intervals within a neighborhood of size $g$ are removed to avoid redundant detections. The procedure continues until no intervals remain. The pseudo-code is summarized in Algorithm S1 in the Supplementary Material. Since the seeded interval design ensures deterministic coverage of the search space while maintaining computational efficiency, it is particularly suitable in our setting, where evaluating the test statistic requires solving an optimization problem.}

\textcolor{black}{While such search strategies are well developed in the change-point literature, their performance in our setting depends critically on the choice of the detection threshold $\zeta_n$. According to Theorem \ref{thm:searching}, the optimal threshold involves unknown quantities such as derivatives of the latent parameter $\beta$, making it difficult to specify directly. To address this issue, we introduce a data-driven screening procedure that identifies a candidate set of structural changes before applying a formal multiple-testing step.}

\textcolor{black}{The screening procedure aims to remove intervals that are unlikely to contain structural changes, which we refer to as null intervals. A key observation is that the seeded interval construction provides structural information about how many intervals can remain unaffected by true changes. Leveraging this property, we derive a lower bound for the number of null intervals when an upper bound $K_{\max}$ on the number of structural changes is specified.}

\begin{prop}
\label{prop:number-of-no-change}
Assume that the true number of null intervals is $N(K_n)$, and $\hat{N}(K_{\max})$ is the number of null intervals when $K_{\max}$ structural changes are equally spaced. If $K_{\max} \geq 5K_n/4$, then $N(K_n) > \hat{N}(K_{\max})$. When the decay parameter $a = 1/2$, the condition can be reduced to $K_{\max} \geq K_n$.
\end{prop}

Proposition \ref{prop:number-of-no-change} shows that if the number of structural changes is moderately overestimated, we can obtain a valid lower bound for the number of null intervals. Based on this, we define a screening threshold  $\tilde{\zeta}_n$ as the $1 - \hat{N}(K_{\max})$ quantile of the maximum contrasts $\left\{ \Delta_{(s, e]}: (s, e] \in \mathcal{I} \right\}$ for screening. Let $\tilde{\mathcal{T}} = \left\{ \tilde{\tau}_1, \dots, \tilde{\tau}_{\tilde{K}_n} \right\}$ denote the collection of candidate structural changes after screening. We then apply the FDR control procedure in Section \ref{subsec:FDR} to obtain the final set of estimated structural changes. 

\textcolor{black}{The proposed screening procedure is an important component of our methodology. Rather than relying on a carefully calibrated detection threshold within the search algorithm, the screening step provides a principled way to construct a manageable set of candidate change points prior to formal multiple testing. Importantly, unlike approaches that select the top $K_{\max}$ structural changes along a solution path, our procedure adapts to the empirical distribution of the contrast statistics and therefore determines the candidate set in a data-driven manner. This avoids committing to a fixed number of candidates and improves robustness when the true number of structural changes is unknown. Moreover, the screening principle relies only on the structural properties of seeded intervals and the availability of a contrast statistic; it provides a general and computationally efficient mechanism for candidate generation in change-point detection problems that employ seeded interval designs.}

\subsection{FDR control procedure}
\label{subsec:FDR}
Before introducing our procedure for controlling the false discovery rate (FDR), we first define what constitutes a false discovery in the context of multiple structural change detection. Specifically, we adopt a definition following \cite{Munk16}: a candidate \( \tilde{\tau}_j \) is considered informative if there exists a true structural change within the interval $\left[ \lceil (\tilde{\tau}_{j-1} + \tilde{\tau}_j)/2 \rceil, \lceil (\tilde{\tau}_{j} + \tilde{\tau}_{j+1})/2 \rceil \right)$, where $\tilde{\tau}_0 = 0$ and $\tilde{\tau}_{K_{\max}+1} = n$; Otherwise, \( \tilde{\tau}_j \) is classified as uninformative. Let \( \mathcal{J}_1(\tilde{\mathcal{T}}) \) and \( \mathcal{J}_0(\tilde{\mathcal{T}}) \) denote the sets of informative and uninformative candidates, respectively. The false discovery proportion (FDP) is then defined as
\[
\mathrm{FDP}(\tilde{\mathcal{T}}) = \frac{\#\left\{ j: \tilde{\tau}_j \in \mathcal{J}_0(\tilde{\mathcal{T}}) \right\}}{\max(|\tilde{\mathcal{T}}|, 1)},
\]
and the FDR is its expectation. \textcolor{black}{More details about the FDR definition are deferred to Section S2 of the Supplementary Material.}

We now introduce our FDR control method, the \textit{Symmetric Parameter Comparison (SPC)} procedure. Let the data be evenly divided into odd and even-indexed subsamples: $\mathcal{Z}_O = \{ y(t_{2i-1}) : i = 1, \dots, n/2 \}$, $\mathcal{Z}_E = \{ y(t_{2i}) : i = 1, \dots, n/2 \}$
assuming \( n \) is even for simplicity. This split preserves the underlying change structure while ensuring minimal discrepancy between the two samples.

We apply the SNOT algorithm to \( \mathcal{Z}_O \) to obtain the candidate set \( \tilde{\mathcal{T}} = \{ \tilde{\tau}_1, \dots, \tilde{\tau}_{\tilde{K}_n} \} \). For each \( \tilde{\tau}_j \), we compute a ranking statistic to assess its credibility as a true structural change:
\begin{equation}
\label{eq:MOPS_statistic}
W_j = \frac{n_j n_{j+1}}{n_j + n_{j+1}} \left( \tilde{\beta}_j^O - \tilde{\beta}_{j+1}^O \right)^\top \Omega_n \left( \tilde{\beta}_j^E - \tilde{\beta}_{j+1}^E \right),
\end{equation}
where \( n_j = \tilde{\tau}_j - \tilde{\tau}_{j-1} \), and \( \tilde{\beta}_j^O \), \( \tilde{\beta}_j^E \) are the estimators of \( \beta \) in the \( j \)th segment using odd and even subsamples, respectively. Here $\Omega_n$ serves as a rough scale estimator for parameters, and the choice of $\Omega_n$ is not crucial.

Intuitively, large and positive values of \( W_j \) suggest strong evidence of change, while large negative values are indicative of no change happening. Asymptotically, for uninformative candidates \( \tilde{\tau}_j \in \mathcal{J}_0 \), the distribution of \( W_j \) is symmetric about zero. This motivates the symmetry-based thresholding:
\begin{equation}
\label{eq:symmetric}
\sup_{0 \leq t < M_n} \left| \frac{\sum_{\tilde{\tau}_j \in \mathcal{J}_0} \mathbb{I}(W_j \geq t)}{\sum_{\tilde{\tau}_j \in \mathcal{J}_0} \mathbb{I}(W_j \leq -t)} - 1 \right| = o_p(1),
\end{equation}
for some large \( M_n \), under mild regularity conditions.

Then the data-driven threshold \( L > 0 \) is selected by
\begin{equation}
\label{eq:FDR-thres}
L = \inf \left\{ t > 0 : \frac{1 + \#\{ j : W_j \leq -t \}}{\#\{ j : W_j \geq t \}} \leq \alpha \right\},
\end{equation}
where \( \alpha \in (0, 1) \) is the target FDR level. Then the final set of selected informative structural changes is $\mathcal{M} = \{ \tilde{\tau}_j : W_j \geq L \}.$ If no such \( W_j \) exceeds \( L \), we set \( L = +\infty \) and \( \mathcal{M} = \emptyset \).
The rationale for this rule is as follows. The numerator in \eqref{eq:FDR-thres} overestimates the number of false positives by symmetry due to \eqref{eq:symmetric}, and the denominator counts total selections. As a result, the ratio is a conservative estimator of the FDP, ensuring valid FDR control. In practice, the empirical FDR is typically close to the nominal level \( \alpha \), because most large negative \( W_j \) values arise from uninformative candidates. Moreover, this approach bypasses the need for asymptotic or simulated null distributions, enabling flexible and accurate thresholding entirely based on data.

\subsection{Generalization to growth systems}
\label{subsec:growth}
The previous test statistics are typically built for finite dynamic systems whose underlying signal is bounded in the time domain. It includes a wide range of systems, such as many periodic models (e.g., the FitzHugh–Nagumo equation), steady-state systems (e.g., energy balance equation), and certainly, most of the systems defined on compact time domains.

The growth system, whose time domain is unbounded and the underlying signal grows unboundedly with time, i.e., $\abs{u(t; \beta)} \to \infty$ as $t \to \infty$, is also important in the analysis of dynamic systems. \textcolor{black}{In such systems, structural change detection entails \textit{regime-wise difficulty}. This includes diverging initial-value components, for which estimation may involve parameters increasing with $n$, and scaling incompatibility, because different parameters may evolve on different natural scales}. We elaborate these two issues in detail in the case of the classical linear model $u(t; \theta, \gamma) = \gamma + \theta t$, see Example \ref{example:linear}, and then provide necessary modifications on $\Delta_{(s, e]}(r)$ and $W_j$ to ensure the applicability of our proposed procedure.
\begin{exam}
    \label{example:linear}
    Let \(u(t;\theta,\gamma)=\gamma+\theta t\) be observed at integer times \(t_i=i\) for \(i=1,\dots,n\).
Assume a single change at \(\tau_1\) (with \(\tau_1\to\infty\) as \(n\to\infty\)):
\[
u(i) = \theta_0 i\cdot\mathbf{1}\left\{1\le i\le\tau_1\right\} + (2\theta_0\tau_1-\theta_0 i)\cdot \mathbf{1}\left\{\tau_1< i\le\tau_2\right\},   \theta_0>0,
\]
and let \(y_i=u(i)+\varepsilon_i,\ \varepsilon_i\stackrel{\text{i.i.d.}}{\sim}\mathcal{N}(0,1)\).
Separate least-squares fits on \([1,\tau_1]\) and \([\tau_1+1,\tau_2]\) reveal that the true intercept in the second segment,
\(\gamma_2=2\theta_0\tau_1\), diverges linearly. While this specific system admits a closed-form solution, in general growth systems, such diverging initial values may cause instability or inconsistency in parameter estimation. More critically, the variances of the intercept and slope estimators differ in scale, reflecting scaling incompatibility. For example, let $\left(\hat{\gamma}_1, \hat{\theta}_1\right)^\top = \arg\min_{\gamma, \theta}\sum_{i=1}^{\tau_1}\left( y_i - \gamma - \theta i \right)^2$, then $\operatorname{Var}(\hat{\gamma}_1) \sim \tau_1^2/\tau_1^3$ and $\operatorname{Var}(\hat{\theta}_1) \sim 1/\tau_1^3$.
\end{exam}
To accommodate such heterogeneity, we introduce segment-specific diagonal scaling matrices.  
For an interval \((s,e]\) we use 
\(
A(s,e)=\mathrm{diag}\bigl\{ \sqrt{(e-s)^3}/(e+s),\; \sqrt{(e-s)^3} \bigr\},
\)
which normalizes intercept and slope on their natural rates.  
In the statistics of Sections \ref{subsec:statistics}--\ref{subsec:FDR}, we simply replace each raw contrast by its scaled version:
\[
\Delta_{(s,e]}(r)=\bigl\|F(s,e,r)\bigl(\hat\beta_{(s,r]}-\hat\beta_{(r,e]}\bigr)\bigr\|_2,\qquad
W_j=\left\{ F_{j}\bigl(\tilde\beta_j^{O}-\tilde\beta_{j+1}^{O}\bigr) \bigr)^{\!\top}\!
           \Omega_n\,
           \bigl( F_{j}\bigl(\tilde\beta_j^{E}-\tilde\beta_{j+1}^{E}\bigr) \right\},
\]
where \(F(s,e,r)=A(r,e)\{A^2(s,r)+A^2(r,e)\}^{-1/2}A(s,r)\) and 
\(F_{j}=F(\tilde\tau_{j-1},\tilde\tau_{j+1},\tilde\tau_{j})\).
Although shown here for the linear model, exactly the same construction applies whenever the signal can be separated into an initial-value term and a diverging dynamic term to avoid the influence of infinite initial values.

Formally, we call a system a {\it separable growth system} if its solution can be written as 
\(
u(t;\theta,\gamma)=\gamma\varphi(t)+\psi(t;\theta),
\)
with \(\varphi\) continuous and uniformly bounded on \((0,\infty)\), 
\(\psi\) continuous in \((t,\theta)\) and diverging as \(t\to\infty\), and \(\varphi(0)=1,\ \psi(0;\theta)=0\).
For such systems, the matrix \(A(s,e)\) can always be chosen to satisfy the mild regularity condition stated in Assumption S1 of the Supplementary Material, thereby guaranteeing that the scaled contrasts remain asymptotically normal.

This structural separation resolves the issue of infinite initial values. It ensures that the diverging behavior of the system is attributed solely to the systematic component $\psi(t;\theta)$, while the intercept-like parameter $\gamma$ governs only the bounded term $\varphi(t)$. Consequently, inference on $\theta$ remains unaffected by the divergence of $\gamma$, as the variance of $\hat{\theta}$ becomes independent of $\gamma$.
\begin{Rmk}
\label{rmk:vanish}
For the intercept in Example \ref{example:linear} the factor
\(\sqrt{(e-s)^3}/(e+s)\) tends to zero as the segment moves farther from the origin; nevertheless, the scaled intercept contrast can still diverge when two adjacent regimes possess intercepts that themselves grow without bound. Indeed, with
\(a_1=\sqrt{\tau_1}\) and \(a_2=\sqrt{(\tau_2-\tau_1)^3}/(\tau_2+\tau_1)\),
\[
\frac{a_1a_2}{\sqrt{a_1^2+a_2^2}}\bigl(\gamma_1-\gamma_2\bigr)
       \asymp \sqrt{\frac{(\tau_2-\tau_1)^3\tau_1^3}{(\tau_2+\tau_1)^2\tau_1}}
       \gtrsim \delta_n^{3/2},
\]
so the contrast dominates \(\sqrt{\log n}\) provided \(\delta_n^{3/2}\gtrsim\sqrt{\log n}\).  
Hence even a shrinking normalizer yields a test statistic that diverges under the alternative and thus preserves power.
\end{Rmk}
\textcolor{black}{In summary, the above generalization extends the proposed methodology beyond finite systems to those exhibiting unbounded growth. By introducing a separable representation of the solution and appropriate segment-specific scaling, the estimation and testing procedures remain valid even when the trajectories diverge. The same construction also accommodates heterogeneous regimes in which different components of the system correspond to initial-value effects, long-run dynamic effects, or distinct estimation scales. Consequently, the proposed approach constitutes a unified structural change detection framework applicable to both finite and diverging dynamical systems.}
\section{Theoretical Analysis}
\label{sec:thm}
In this section, we present the non-asymptotic error bounds of the parameter estimators and the detection properties of the SNOT searching algorithm based on the proposed test statistics. We then establish guarantees for our FDR control procedure. For notational simplicity, all theoretical results are stated for the univariate case $d = 1$; the extension to higher dimensions is straightforward. For space economy, the theoretical analysis for separable growth systems is deferred to Section S3 of the Supplementary Material.

\subsection{Detection Property}
Before presenting the detection properties, we introduce several standard assumptions on the ODE model to ensure the validity of parameter estimation. 

\begin{assum}
\label{ass:ODE}
We assume that the dynamic system $u(\cdot;\beta)$ and the sampling points $t_1, \dots, t_n$ satisfy the following conditions,\\
\noindent (a) (Identifiability) For any $\beta_1 \neq \beta_2$, $\sum_{i=s + 1}^{e}\left\{u(t_i;\beta_1) - u(t_i;\beta_2)\right\}^2/(e -s) > 0$ for any $e - s > p+1$.\\
\noindent (b) (Continuity) For $i = 1, \dots, n$, the first and second derivatives of $u(t_i;\beta)$ with respect to $\beta$ are continuous over a compact parameter space $\Theta \times \mathcal{U}$.\\
\noindent (c) (Stability) Define the Gram matrices as $G(\beta, s, e) = \sum_{i=s + 1}^{e}\frac{\partial u(t_i;\beta)}{\partial\beta}\frac{\partial u(t_i;\beta)}{\partial\beta^\top}/(e - s)$, we assume that the minimum and maximum eigenvalue of $G(\beta, s, e)$ satisfies that $\phi_0 \leq \phi_{\min}(G(\beta, s, e)) \leq \phi_{\max}(G(\beta, s, e)) \leq \phi_1$ uniformly for $e - s > p + 1$, and $\beta \in \Theta \times \mathcal{U}$.\\
\noindent (d) (Lipschitz) There exists constants $L_{\nabla}$ and $L_{H}$ such that for $i = 1, \dots, n$, $\norm{\frac{\partial u(t_i;\beta)}{\partial\beta}}_2 \leq L_{\nabla}$, $\norm{\frac{\partial^2 u(t_i;\beta)}{\partial\beta\beta^\top}}_{\mathrm{op}} \leq L_H$ uniformly for $\beta \in \Theta \times \mathcal{U}$.\\
\noindent (e) The numerical solution $\hat{u}(t;\beta)$ of the ODE satisfies that there exists a predetermined numerical error tolerance $\epsilon_u = o(1)$ such that $\sum_{i=1}^n\left\{u(t_i;\beta) - \hat{u}(t_i;\beta)\right\}^2 < \epsilon_u$ uniformly for $\beta \in \Theta\times\mathcal{U}$.
\end{assum}
 
Conditions (a)–(c) are standard assumptions used to establish the consistency and asymptotic normality of the least squares estimator in inverse problems involving ordinary or partial differential equations \citep{Cao07,chen2025deep}, and have also been extensively studied in the context of nonlinear regression \citep{Seber03}. Condition (d) imposes Lipschitz continuity on the first and second derivatives with respect to $\beta$. Although this requirement is slightly stronger than what is typically needed for parameter estimation, it is essential in the structural change detection setting, where we must control the tail behavior of the estimator $\hat{\beta}$ at the scale of $\sqrt{\log n}$. Condition (e) ensures that the numerical solution is sufficiently accurate. \textcolor{black}{Under the bi-Lipschitz conditions (c)(d), modern ODE solvers satisfy that, with adaptive step-size control, the numerical error can be controlled through prescribed tolerances, $\epsilon_u = o(1)$, and made negligible relative to the statistical error, even in the case that $t_n \rightarrow \infty$. For rigorous proofs, see Section S4 in the Supplementary Material.} Under this assumption, we can derive the following concentration bound for the parameter estimators.

\begin{thm}
\label{thm:estimation}
Suppose that we have $n$ observations $y(t_1), \dots, y(t_n)$ from the dynamic system \eqref{eq:ode} (i.e., without structural change) and Assumption \ref{ass:ODE} holds. Denote the true parameter as $\beta_0$ and then, with probability at least $1 - c_3\exp(-c_4n)$, 
\begin{equation*}
\label{eq:M-tail}
\mathbb{P}\left\{ \sqrt{n}\norm{\hat{\beta} - \beta_0}_2 \geq \rho + o(1) \right\} \leq 2\exp\left\{ -c_1\min\left( \frac{\phi_0^2\rho^2}{\sigma^2L^2_{\nabla}}, \frac{\sqrt{n}\phi_0\rho}{\sigma L_{\nabla}} \right) \right\},
\end{equation*}
and
\begin{equation*}
\mathbb{P}\left\{ \sqrt{n}\norm{\hat{\beta} - \beta_0}_2 \geq \rho - o(1) \right\} \geq 2\exp\left( -\frac{9c_1\rho^2}{\sigma^2} \right) - \frac{c_2}{\sqrt{n}}, 
\end{equation*}
where $c_1, \ldots, c_4 > 0$ are constants, and we remark here that $L_{\nabla} \geq \phi_1 \geq \phi_0$.
\end{thm}

This theorem establishes the theoretical foundation for our detection procedure by providing sharp, two-sided non-asymptotic bounds on the estimation error. With a desired numerical precision, the contribution of the numerical approximation becomes asymptotically negligible, ensuring that the numerical solution does not affect either the estimation accuracy or the tail probability of the estimator. The lower and upper bounds together yield precise control over the stochastic fluctuations of $\hat{\beta}$, guaranteeing both accurate identification of structural changes and valid FDR control in the subsequent detection procedure. We note that such probabilistic bounds possibly cannot be guaranteed for methods relying on nonparametric smoothing or trajectory reconstruction; as a result, these approaches face substantial difficulty in constructing validated searching and FDR control procedures.


For the dynamic system \eqref{eq:ode_change} which has structural changes, we impose an assumption on the gaps between structural change locations and the magnitudes of changes.  

\begin{assum}
    \label{ass:gaps and magnitude}
    Let $\delta_j = \tau_j^* - \tau_{j-1}^{*}$, $j = 1, \dots, K_n + 1$. The minimum spacing $\delta_n = \min_{j=1, \dots, K_n + 1}\delta_j$ and the minimum magnitudes of jumps $\underline{\beta}_n = \min_{j = 1, \dots, K_n + 1}\Delta_{j}^{\beta} = \min_{j = 1, \dots, K_n + 1}\norm{\beta^*_{j} - \beta^*_{j-1}}_2$, satisfy that $\delta_n^{1/2}\underline{\beta}_n \geq C_1 \sqrt{\log(n)}$ for a certain constant $C_1 > 0$.
\end{assum}

Assumption \ref{ass:gaps and magnitude} establishes a relationship between the minimum spacing of structural changes and the minimum jump magnitude with the quantity $\delta_n^{1/2}\underline{\beta}_n$, which determines the number and distributions of the structural changes. Intuitively, the inequality $\delta_n^{1/2}\underline{\beta}_n \geq C_1 \sqrt{\log(n)}$ implies that a large magnitude of jumps allows for identifying more changes, and we need relatively more samples for detecting small jumps. Under Assumptions \ref{ass:ODE}  and \ref{ass:gaps and magnitude}, we obtain the following result.

\begin{thm}
\label{thm:searching}
Suppose that we have $n$ observations $y(t_1), \dots, y(t_n)$ from the dynamic system \eqref{eq:ode_change} and Assumptions \ref{ass:ODE} and \ref{ass:gaps and magnitude} hold. 
Let $\hat{K}_n$ and $\hat{\tau}_1, \dots, \hat{\tau}_{\hat{K}_n}$ denote, respectively, the number and locations of the estimated structural changes. There exists two constants $C_2 = C_1/2a^2(1 - a^2), C_3 = 1/2a^2(1 - a^2)$ such that if $C_2\sqrt{\log n} \leq \zeta_n < C_3\delta_n^{1/2}\underline{\beta}_n$ and $m \leq a^2\delta_n$, as $n \rightarrow \infty$,
\begin{equation*}
\mathbb{P}\left\{ \hat{K}_n = K_n, \max_{j = 1, \dots, K_n}\abs{\hat{\tau}_j - \tau^*_j}(\Delta_j^\beta)^{2} \leq C_4 \log(n) \right\} \rightarrow 1,
\end{equation*}
for certain constants $C_4 > 0$.
\end{thm}

Theorem \ref{thm:searching} established the detection consistency of the SNOT searching algorithm. In contrast to Theorem 1 in \cite{Baranowski19}, this consistency does not depends on the total number of seeded intervals $M$, but solely on their minimum length $m$. Indeed, when $m \le a^{2}\delta_{n}$, every true structural change $\tau^{*}_{j}$ lies in a seeded interval $(s_{j}, e_{j}]$ whose length $\ell_{j}=e_{j}-s_{j}$ satisfies $\ell_{j}>2a^{2}\delta_{n}$ and whose midpoint $r_{j}$ obeys $|r_{j}-\tau^{*}_{j}| \le \ell_{j}(1+a^{2})/4$. Such an interval is therefore long enough for the proposed statistic to identify the unique structural change it contains. Furthermore, if $m$ is fixed, the collection $\mathcal{I}$ contains $O(n)$ seeded intervals whose total length is $O(n\log n)$; consequently, the overall computational cost of SNOT is $O(n\log n)$. By contrast, methods based on random background intervals incur a cost of order $n^3\log(n^2/\delta_n)/\delta_n^2$. Hence, whenever $\delta_{n}=o(n)$, the SNOT algorithm is more computationally efficient than methods with random background intervals. 

Moreover, this theorem allows for $\delta_n^{1/2}\underline{\beta}_n$, which captures the intrinsic difficulty of the detection problem, to be as small as order $\sqrt{\log n}$. As argued in \cite{Chan13}, this is the minimal rate required for consistent change point detection from a minimax perspective. The bound $\max_{j=1, \dots, K_n} |\hat{\tau}_j - \tau^*_j| (\Delta_j^\beta)^{2} \leq C_4 \log n$ quantifies the maximum discrepancy between the estimated and true structural change locations, weighted by the squared jump magnitudes. This implies that the minimal signal strength cannot be arbitrarily weak, which is reasonable as it serves as the common identifiability condition. For $\underline{\beta}_n \sim 1$, the maximum spacing is of order $O_p(\log n)$, which trails the minimax rate of $O_p(1)$ by only a logarithmic factor.

\begin{Rmk}
   \textcolor{black}{For the classical separable growth systems--linear system, the theoretical properties in Section S3 of the Supplementary Material illustrate that under the same signal-strength condition as in \cite{Baranowski19}, the proposed framework yields a localization error of order
$$\min_j\left\{ \left( \Delta_j^{\theta} \right)^{-2/3}\log^{1/3}(n), \left( \Delta_j^{\gamma}/\tau_j^* \right)^{-2/3}\log^{1/3}(n) \right\}.$$
The result in \cite{Baranowski19} is recovered through the first term, corresponding to the slope-change component. The additional term shows that our method also accounts for localization driven by intercept changes, thereby providing a strictly more general characterization in the piecewise linear setting.} 
\end{Rmk}

As a consequence, the candidate set of structural changes $\tilde{\mathcal{T}}$, obtained via the threshold $\tilde{\zeta}_n$, satisfies the following consistency property:
\begin{equation}
\label{eq:consistent cover}
    \mathbb{P}\left\{ \tilde{K}_n \geq K_n,\max_{j = 1, \dots, K_n}\abs{\hat{\tau}_j - \tau^*_j}(\Delta_j^\beta)^{2}\leq C_4 \log(n) \right\} \rightarrow 1, 
\end{equation}
where $\hat{\tau}_j = \arg\min_{\tilde{\tau}_k \in \tilde{\mathcal{T}}}\abs{\tau_k - \tau^*_j}$. This result implies that $\tilde{\mathcal{T}}$ provides a consistent cover of the true structural changes, capturing all structural changes with high probability. Building on this result, we now proceed to establish the theoretical guarantees for the FDR control procedure. In particular, we show that the final set of structural change estimates, obtained by applying the SPC procedure to $\tilde{\mathcal{T}}$, achieves valid false discovery rate control.

\subsection{FDR control property}

Now, we focus on the asymptotic property of the proposed FDR control procedure. Recall that the candidate set of structural changes is denoted by $\tilde{\mathcal{T}} = \{ \tilde{\tau}_1, \dots, \tilde{\tau}_{\tilde{K}_n} \}$, with $\mathcal{J}_0(\tilde{\mathcal{T}})$ and $\mathcal{J}_1(\tilde{\mathcal{T}})$ representing the sets of uninformative and informative candidates, respectively. For notational convenience, we denote these sets simply as $\mathcal{J}_0$ and $\mathcal{J}_1$ in what follows. We impose a theoretical requirement for the spacings among the structural changes in $\tilde{\mathcal{T}}$. 
\begin{assum}
\label{ass:minimum distance}
Let $\mathcal{T}(\omega_n) = \left\{ \mathcal{T}: \min_{j}(\tau_{j+1} - \tau_j) \geq \omega_n \right\}$, then $\tilde{\mathcal{T}} \subset \mathcal{T}(\omega_n)$ with $\omega_n < \delta_n$. 
\end{assum}
Assumption \ref{ass:minimum distance} imposes a lower bound on the minimal separation between estimated structural changes. This is necessary because inference on estimated change points requires a sufficient amount of sample information within each segment. Empirically, this assumption can be satisfied by appropriately choosing the minimal gap length $g$ (e.g., $g = m/2$) in the searching Algorithm S1.

In addition, to ensure the validity of the FDR control procedure, it is reasonable to require that each true structural change is closely approximated by at least one estimated candidate. Specifically, we require that the candidate structural changes satisfy that $\tilde{K}_n \geq K_n$ and there exist $\tilde{\tau}_{j_1}< \cdots < \tilde{\tau}_{j_{K_n}} \in \tilde{\mathcal{T}}$ such that $\abs{\tilde{\tau}_{j_k} - \tau^*_{k}} \leq \lambda_{nk}$ holds with probability approaching one as $n \rightarrow \infty$. According to Theorem \ref{thm:searching} and equation \eqref{eq:consistent cover}, this condition is automatically satisfied with $\lambda_{nk} = C_4(\Delta_k^\beta)^{-2}\log n$. Under this setup, we establish the following FDR control result.


\begin{thm}
\label{thm:FDR}
Suppose Assumptions \ref{ass:ODE}--\ref{ass:minimum distance} hold, then if $\omega_n^{1/2}\underline{\beta}_n/\log n \rightarrow \infty$ as $n \rightarrow \infty$, we have that $\lim_{n\rightarrow \infty} \mathbb{P}\left( \mathcal{J}_1 \subset \mathcal{M} \right) = 1$, and for any $\alpha \in (0, 1)$, $\mathrm{FDP}(\mathcal{M}) \leq \alpha + o_p(1)$. It follows that $\lim\sup_{n \rightarrow \infty} \mathrm{FDR}(\mathcal{M}) \leq \alpha$.
\end{thm}

This theorem implies that, under an identifiable signal strength, the FDR control procedure can consistently detect all structural changes while asymptotically controlling the FDR level. The condition $\omega_n^{1/2} \underline{\beta}_n / \log n \to \infty$ implies a wider spacing between true structural changes than required by Assumption \ref{ass:gaps and magnitude}. However, the overall selection procedure still achieves consistent recovery even when the minimum spacing $\delta_n$ between true structural changes is only of order $\log^k(n)$ for some $k > 2$.

Compared to the theoretical guarantees of the strengthened Schwarz Information Criterion (sSIC; see Theorem 3 in \cite{Baranowski19}), our SPC procedure requires slightly larger spacing when $\underline{\beta}_n \sim 1$, as sSIC needs $\omega_n / \log^{\alpha'}(n) \to \infty$ for some $\alpha' > 1$. However, unlike sSIC, our approach does not require the number of structural changes $K_n$ to remain fixed, making the SPC procedure more broadly applicable in practice. Furthermore, empirical results reported in Table \ref{table:Linear} suggest that the sSIC procedure tends to be relatively conservative, as the difference in the numerical performance between the proposed test statistic and the GLR statistic is likely to be minimal.

\section{Simulation Studies}
\label{sec:simulation}

In this section, we conduct simulation studies to evaluate the numerical performance of our proposed method and existing methods in detecting structural changes in complex dynamic systems. For fair comparison, we also conduct simulations in linear and linear regression systems to illustrate that the proposed method is broadly applicable, where the linear system is also a typical separable growth system.

\subsection{FitzHugh–Nagumo Equation Model}
The dynamic system we chose here is the FitzHugh–Nagumo (FN) equation, which was developed by \cite{Nagumo62} to model the behavior of spike potentials in the giant axon of squid neurons:
\begin{equation*}
\frac{dV}{dt} = 3(V - \frac{V^3}{3} + R),\quad \frac{dR}{dt}  = -\frac{1}{3}(V - 0.2+ \theta R).
\end{equation*}
The system is governed by the systematic parameter $\theta$, along with the initial values $V_0 = V(0)$ and $R_0 = R(0)$. We denote this system as $FN(t; b, V_0, R_0)$, and use $DFN(t)$ to represent the entire system when structural changes are present. To illustrate our method, we consider two representative scenarios involving structural changes. Recall that there are $K_n$ structural changes located at time points $\tau^*_1, \dots, \tau^*_{K_n}$. For notational convenience, we let $\tau^*_0 = 1$ and $\tau^*_{K_n+1} = n$. The tested models are then defined as follows. \\
(1) \textit{General structural change}. In the initial segment $[\tau^*_0, \tau^*_1]$, the system is defined as $FN(t; \theta_0, -1, 1)$. For each subsequent segment $(\tau^*_j, \tau^*_{j+1}]$ with $j = 1, \dots, K_n$, we assign a distinct systematic parameter $\theta_j$. The start time for the system on each segment is set to $t_{\tau^*_j}$, and the corresponding initial value is given by $DFN(t_{\tau^*_j}) + \mu_j$, where $\mu_j$ introduces a shift in initial condition to reflect general structural change. \\
(2) \textit{Systematic structural change}. This setting follows the same construction as in (1), but the initial value at time $t_{\tau^*_j}$ is taken directly as $DFN(t_{\tau^*_j})$, without any shift. This ensures continuity of the trajectory across structural changes.

The first setting represents general structural changes that may occur in various scientific applications. The second, systematic structural change, is more typical in social or epidemiological systems governed by evolving but policy-driven regimes, where the dynamic trajectory remains continuous across structural changes. A representative example includes modeling the spread of COVID-19, where distinct policy phases lead to changes in transmission dynamics without abrupt changes in observed variables.

We set $n = 2000$ and $K_n = 10$, with observation times defined as $t_j = 0.1 \times j$ for $j = 1, \dots, n$. The location of the $k$-th structural change is given by $\tau^*_k = j \lfloor n / (K_n + 1) \rfloor + \mathrm{Unif}(-n^{1/4}, n^{1/4})$, for $j = 1, \dots, K_n + 1$, where $\mathrm{Unif}(a, b)$ denotes the uniform distribution over the interval $[a, b]$. The full set of $n$ observations is generated from the complete system $DFN(t)$ with added noise, specifically, $y(t_i) = DFN(t_i) + \varepsilon_i, i = 1, \dots, n$, where each noise term $\varepsilon_i = (\varepsilon_{i1}, \varepsilon_{i2})^\top$, and the components $\varepsilon_{ij}$, $j = 1, 2$, are independently and identically distributed according to either a normal distribution $\mathcal{N}(0, \sigma^2)$ or a $t$-distribution $\sigma t(\mathrm{3})/\sqrt{3}$.

The systematic parameters $\theta_j$, for $j = 0, \dots, K_n$, are drawn independently from $\mathcal{N}(0.6, 0.2)$. For general structural change scenarios, the perturbations $\mu_j$ are sampled from $\mathcal{N}(0, 0.2)$. To control the noise-to-signal ratio (NSR), the noise standard deviation is defined as $\sigma = \gamma \underline{u}_n$, where
\begin{equation*}
\label{eq:signalhat}
    \underline{u}_n = \min_{j=1, \dots, K_n}\sqrt{\frac{1}{n_j + n_{j+1}}\sum_{i = {\tau_{j}^*}}^{{\tau_{j+1}^*}}\left\{ V(t_i;\beta_{j+1}^*) - V(t_i;\beta_j^*)\right\}^2 + \left\{R(t_i;\beta_{j+1}^*) - R(t_i;\beta_j^*) \right\}^2},
\end{equation*}
and $\gamma \in \left\{ 0.5, 0.7, 0.9, 1.1, 1.3 \right\}$ specifies the target NSR level. Further discussion on the interpretation and computation of $\underline{u}_n$ is provided in Section S5 of the Supplementary Material.

The performance of the proposed method is evaluated using the false discovery rate (FDR) along with several additional indices. These include the true positive rate (TPR), defined as the proportion of informative locations that are correctly identified, and the average number of estimated structural changes, denoted by $\hat{K}$. To assess the accuracy of the estimated structural change locations, we also report the distance between the estimated set $\mathcal{M}$ and the true structural change set $\mathcal{T}^*$, defined as $d(\mathcal{M}, \mathcal{T}^*) =  \sum_{\tau_k^* \in \mathcal{T}^*} \min_{\hat{\tau}_j \in \mathcal{M}} |\hat{\tau}_j - \tau_k^*|/|\mathcal{T}^*|$, which reflects the average minimum deviation of estimated locations from the true structural change locations. The detection results for our proposed method under both Model (1) and Model (2) are summarized in Table \ref{table:FN}.

\begin{table}[h]
  \setlength{\abovecaptionskip}{0cm}
  \caption{Detection results for system $DFN$ with general and systematic structural changes. \label{table:FN}} 
  \setlength{\tabcolsep}{3mm}
  \centering
   \resizebox{1\textwidth}{1.4in}{\begin{tabular}{ccccccccc}
    \toprule
    & \multicolumn{4}{c}{Model (1): General} & \multicolumn{4}{c}{Model (2): Systematic}\\
    \cmidrule(r){2-5} \cmidrule(r){6-9}
    NSR($\gamma$) & TPR & FDR & $\hat{K}$ & $d(\mathcal{M}, \mathcal{T}^*)$ & TPR & FDR & $\hat{K}$ & $d(\mathcal{M}, \mathcal{T}^*)$\\
    \midrule
    \multicolumn{9}{c}{Normal distributed error} \\
    0.5 & 0.976 & 0.112 & 11.11 & 7.617 & 0.954 & 0.164 & 11.60 & 23.44\\
    0.7 & 0.958 & 0.107 & 10.87 & 11.81 & 0.942 & 0.170 & 11.56 & 26.45\\
    0.9 & 0.935 & 0.103 & 10.62 & 17.34 & 0.932 & 0.163 & 11.37 & 30.26\\
    1.1 & 0.917 & 0.115 & 10.57 & 23.00 & 0.932 & 0.168 & 11.44 & 32.29\\
    1.3 & 0.883 & 0.126 & 10.37 & 28.86 & 0.906 & 0.154 & 11.00 & 38.98\\
    \midrule
    \multicolumn{9}{c}{t distributed error, $\mathrm{df} = 3$}\\
    0.5 & 0.981 & 0.129 & 11.42 & 7.736 & 0.954 & 0.165 & 11.63 & 24.48\\
    0.7 & 0.971 & 0.136 & 11.42 & 11.47 & 0.944 & 0.169 & 11.59 & 28.54\\
    0.9 & 0.946 & 0.139 & 11.23 & 19.29 & 0.927 & 0.178 & 11.48 & 32.76\\
    1.1 & 0.954 & 0.158 & 11.58 & 19.68 & 0.921 & 0.173 & 11.36 & 35.04\\
    1.3 & 0.928 & 0.159 & 11.31 & 24.20 & 0.911 & 0.177 & 11.30 & 36.17\\
    \bottomrule
    \end{tabular}}%
\end{table}

From the results, it is evident that our proposed method consistently achieves accurate FDR control across all scenarios while successfully detecting nearly all structural changes. This empirical finding aligns well with the theoretical guarantees established in Section \ref{sec:thm}. Furthermore, the reported empirical distances demonstrate that the method not only provides a reliable estimate of the number of structural changes but also achieves high precision in localizing their positions.

Despite different model assumptions, we also compare the detection performance in terms of TPR and FDR with the original Narrowest-over-threshold (NOT) method and the Narrowest Significance Pursuit (NSP) method. This comparison is conducted under the setting where only the trajectory $V(t_i) + \varepsilon_{i1}$, for $i = 1, \dots, n$, is observed in Models (1) and (2), as both NOT and NSP are designed for univariate time series. The detection procedures for the NOT and NSP methods are implemented using the R packages {\tt not} and {\tt nsp}. Throughout this study, the final set of estimated structural changes for the NOT method is obtained via the strengthened Schwarz information criterion (sSIC). For NSP, we adopt the midpoints of the estimated intervals as the final estimates. Since the intervals produced by NSP are sufficiently short, the precise choice of representative points has minimal impact on the TPR and FDR calculations. The detection TPRs and FDRs for Model (1) under Gaussian noise are reported in \ref{table:FN-C}.

\begin{table}[h]
  \setlength{\abovecaptionskip}{0.1cm}
  \caption{\small Detection results of different methods for Model (1) (general structural change) with trajectory $V(t)$ only. The notation ``NOT-*'' means that the method is the NOT algorithm with ``*'' contrast function. For example, ``CM'' is the ``pcwsConstMean'' contrast function. One can refer to R package {\tt not} for detailed explanations for all the contrast functions. The notation ``NSP-Poly(*)'' is the NSP algorithm for piecewise-polynomial signals with the polynomial degree ``*''. \label{table:FN-C}} 
  \setlength{\tabcolsep}{2mm}
  \centering
   \resizebox{1\textwidth}{1.2in}{\begin{tabular}{ccccccccccc}
    \toprule
    & \multicolumn{2}{c}{0.5} & \multicolumn{2}{c}{0.7} & \multicolumn{2}{c}{0.9} & \multicolumn{2}{c}{1.1} & \multicolumn{2}{c}{1.3}  \\
    \cmidrule(r){2-3} \cmidrule(r){4-5} \cmidrule(r){6-7} \cmidrule(r){8-9} \cmidrule(r){10-11}
    Method & TPR & FDR & TPR & FDR & TPR & FDR & TPR & FDR & TPR & FDR\\
    \midrule
    Proposed & 0.891  & 0.133  & 0.842  & 0.130  & 0.813  & 0.127  & 0.756  & 0.128  & 0.641  & 0.127  \\
    NOT-CM & 0.921  & 0.460  & 0.933  & 0.469  & 0.934  & 0.467  & 0.943  & 0.471  & 0.951  & 0.470  \\
    NOT-CMHT & 0.938  & 0.476  & 0.942  & 0.470  & 0.944  & 0.471  & 0.953  & 0.478  & 0.954  & 0.474  \\
    NOT-LCM & 0.000  & 0.000  & 0.000  & 0.000  & 0.000  & 0.000  & 0.000  & 0.000  & 0.000  & 0.000  \\
    NOT-LM & 1.000  & 0.491  & 1.000  & 0.491  & 1.000  & 0.491  & 1.000  & 0.491  & 1.000  & 0.491   \\
    NOT-QM & 1.000  & 0.500  & 1.000  & 0.500  & 1.000  & 0.499  & 1.000  & 0.499  & 1.000  & 0.499   \\
    NOT-CMV & 0.966  & 0.492  & 0.981  & 0.492  & 0.973  & 0.493  & 0.971  & 0.493  & 0.968  & 0.493   \\
    NSP-Poly(1) & 1.000  & 0.883  & 1.000  & 0.882  & 1.000  & 0.881  & 1.000  & 0.878  & 1.000  & 0.872   \\
    NSP-Poly(2) & 1.000  & 0.811  & 1.000  & 0.805  & 1.000  & 0.792  & 1.000  & 0.781  & 1.000  & 0.774   \\
    NSP-Poly(3) & 1.000  & 0.876  & 1.000  & 0.872  & 1.000  & 0.870  & 1.000  & 0.869  & 1.000  & 0.867 \\
    \bottomrule
    \end{tabular}}%
\end{table}

From Table \ref{table:FN-C}, it is clear that both the original NOT method and the NSP method perform poorly in detecting structural changes within complex dynamic systems. In contrast, our proposed method maintains effective control over the FDR, and although the TPRs are slightly reduced due to the absence of the trajectory $R(t)$, the results still demonstrate that our method performs reliably even when only a single component of the multivariate system is observed. We also note that the R function ``nsp\_poly\_selfnorm()'', which implements the self-normalized NSP algorithm for piecewise-polynomial signals, occasionally fails when applied to certain realizations of the data. Consequently, we exclude the self-normalized NSP method from our comparison.

\subsection{Linear and Linear Regression Systems}
\label{subsec:linear}

Since the original NOT method and NSP are designed for piecewise polynomial signals, and NSP is also applicable in general linear regression settings, we include additional comparisons to ensure fairness. Specifically, we evaluate the detection performance of our proposed method on the linear system $DL$ and the linear regression system $DLR$, noting that the linear system is a special case of the broader separable growth system, and our method can be directly applied to linear regression. The two models are defined as follows:\\
(3) \textit{Linear system}. We set $t_i = i$, $i = 1, \dots, n$, $DL(1) = 0$ and for $i \in (\tau_j^*, \tau_{j+1}^*]$, $j = 0, \dots, K_n$, $DL(i) = DL(\tau_j^*) + \beta_j(i - \tau_j^*)$, where the systematic parameter $\beta_j$, $j = 0, \dots, K_n$ are sampled from the normal distribution $0.01\times\mathcal{N}(0, 1)$, $n = 2000$ and the number of changes $K_n = 10$. \\
(4) \textit{Linear regression system}. We set $t_i = i$, $i = 1, \dots, n$ and for $i \in (\tau_j^*, \tau_{j+1}^*]$, $j = 0, \dots, K_n$, $DLR(i) = X_i\beta_j$, where the general covariates $X_i, i=1, \dots, n$ are generated from $\mathcal{N}(1, 1)$ and the systematic parameter $\beta_j$, $j = 0, \dots, K_n$ are sampled from $\mathcal{N}(4, 1)$. 

The change magnitude is calculated with the same formula as in equation \eqref{eq:signalhat}, and the NSR parameter $\gamma$ is set to 2.5, 2.7, and 2.9. Table \ref{table:Linear} summarizes the detection performance of various methods applied to the linear system. According to the system design, the appropriate contrast function for the original NOT method is “pcwsLinContMean,” which we refer to as “NOT-LCM.” The polynomial degree for both NSP and the self-normalized NSP is set to 1, denoted respectively as “NSP-Poly(1)” and “NSP-Poly-Self(1).”

The results demonstrate that, across most scenarios, our proposed method achieves the highest true positive rates (TPRs) while maintaining good control of false discovery rates (FDRs). Although the sSIC procedure successfully reduces false positives for NOT-LCM, it tends to be more conservative than our method, as reflected in lower TPRs. NSP-Poly(1) and NSP-Poly-Self(1) produce the lowest FDRs overall, but the former fails to handle heavy-tailed noise in this setting. Moreover, the proposed method has comparable detection accuracy $d(\mathcal{M}, \mathcal{T}^*)$ as NOT-LCM, even if the searching procedure in Section \ref{subsec:algorithm} only involves half of the samples.

\begin{table}[h]
  \setlength{\abovecaptionskip}{0.1cm}
  \caption{Detection results for the linear system $DL$.  \label{table:Linear}} 
  \setlength{\tabcolsep}{2mm}
  \centering
    \resizebox{1\textwidth}{1.7in}{\begin{tabular}{ccccccccc}
    \toprule
    & \multicolumn{4}{c}{Normal distributed error} & \multicolumn{4}{c}{t distributed error}\\
    \cmidrule(r){2-5} \cmidrule(r){6-9}
    Method & TPR & FDR & $\hat{K}$ & $d(\mathcal{M}, \mathcal{T}^*)$ & TPR & FDR & $\hat{K}$ & $d(\mathcal{M}, \mathcal{T}^*)$\\
    \midrule
    \multicolumn{9}{c}{NSR: $\gamma = 2.5$} \\
    Proposed         & 0.947 & 0.194 & 11.95 & 34.97 & 0.963 & 0.203 & 12.25 & 35.71\\
    NOT-LCM          & 0.825 & 0.000 & 8.250 & 36.93 & 0.825 & 0.031 & 8.570 & 37.93\\
    NSP-Poly(1)      & 0.755 & 0.001 & 7.560 & -     & 0.988 & 0.566 & 23.34 & - \\
    NSP-Poly-Self(1) & 0.575 & 0.000 & 5.750 & -     & 0.600 & 0.000 & 6.000 & - \\
    \midrule
    \multicolumn{9}{c}{NSR: $\gamma = 2.7$} \\
    Proposed         & 0.945 & 0.191 & 11.94 & 35.89 & 0.955 & 0.197 & 12.04 & 37.15\\
    NOT-LCM          & 0.818 & 0.001 & 8.190 & 38.32 & 0.824 & 0.022 & 8.450 & 38.51\\
    NSP-Poly(1)      & 0.711 & 0.001 & 7.120 & -     & 0.988 & 0.563 & 23.18 & -\\
    NSP-Poly-Self(1) & 0.559 & 0.000 & 5.590 & -     & 0.574 & 0.000 & 5.740 & -\\
    \midrule
    \multicolumn{9}{c}{NSR: $\gamma = 2.9$} \\
    Proposed         & 0.939 & 0.186 & 11.77 & 38.64 & 0.952 & 0.194 & 11.98 & 37.97\\
    NOT-LCM          & 0.816 & 0.001 & 8.170 & 39.00 & 0.821 & 0.025 & 8.450 & 39.33\\
    NSP-Poly(1)      & 0.692 & 0.001 & 6.930 & -     & 0.989 & 0.561 & 23.11 & - \\
    NSP-Poly-Self(1) & 0.541 & 0.000 & 5.410 & -     & 0.564 & 0.000 & 5.640 & - \\
    \bottomrule
    \end{tabular}}%
\end{table}

We now turn to the detection performance in the linear regression system $DLR$, with results reported in Table \ref{table:Linear-Reg}. Since the original NOT method is not specifically designed for linear regressions with general covariates, we evaluate it using all available candidate contrast functions. Among these, the contrast function “pcwsConstMeanVar” (denoted as NOT-CMV) delivers the best performance in this setting. Therefore, we report only the results of NOT-CMV in the main text, while the performance of other contrast functions is provided in Table S1 of the Supplementary Material. The NSP and self-normalized NSP methods for linear regression with general covariates are denoted as “NSP-Reg” and “NSP-Reg-Self,” respectively.

\begin{table}[h]
  \setlength{\abovecaptionskip}{0.1cm}
  \caption{Detection results for the linear regression system $DLR$.  \label{table:Linear-Reg}} 
  \setlength{\tabcolsep}{2.5mm}
  \centering
    \resizebox{1\textwidth}{1.7in}{\begin{tabular}{ccccccccc}
    \toprule
    & \multicolumn{4}{c}{Normal distributed error} & \multicolumn{4}{c}{t distributed error}\\
    \cmidrule(r){2-5} \cmidrule(r){6-9}
    Method & TPR & FDR & $\hat{K}$ & $d(\mathcal{M}, \mathcal{T}^*)$ & TPR & FDR & $\hat{K}$ & $d(\mathcal{M}, \mathcal{T}^*)$\\
    \midrule
    \multicolumn{9}{c}{NSR: $\gamma = 2.5$} \\
    Proposed     & 0.987 & 0.170 & 12.10 & 9.965 & 0.994 & 0.150 & 11.88 & 9.145\\
    NOT-CMV     & 0.398 & 0.000 & 3.980 & 148.0 & 0.395 & 0.000 & 3.950 & 150.1\\
    NSP-Reg      & 0.832 & 0.001 & 8.330 & -     & 0.973 & 0.544 & 21.98 & -\\
    NSP-Reg-Self & 0.825 & 0.000 & 8.250 & -     & 0.832 & 0.000 & 8.320 & -\\
    \midrule
    \multicolumn{9}{c}{NSR: $\gamma = 2.7$} \\
    Proposed     & 0.983 & 0.167 & 12.01 & 10.83 & 0.989 & 0.151 & 11.83 & 10.29\\
    NOT-CMV     & 0.396 & 0.000 & 3.960 & 149.4 & 0.392 & 0.000 & 3.920 & 152.1\\
    NSP-Reg      & 0.822 & 0.001 & 8.220 & -     & 0.967 & 0.545 & 21.91 & -\\
    NSP-Reg-Self & 0.816 & 0.000 & 8.160 & -     & 0.821 & 0.000 & 8.210 & -\\
    \midrule
    \multicolumn{9}{c}{NSR: $\gamma = 2.9$} \\
    Proposed     & 0.976 & 0.158 & 11.82 & 12.39 & 0.982 & 0.150 & 11.75 & 11.67\\
    NOT-CMV     & 0.395 & 0.000 & 3.950 & 150.1 & 0.391 & 0.000 & 3.910 & 152.8\\
    NSP-Reg      & 0.815 & 0.001 & 8.150 & -     & 0.964 & 0.547 & 21.91 & -\\
    NSP-Reg-Self & 0.811 & 0.000 & 8.110 & -     & 0.817 & 0.000 & 8.170 & -\\
    \bottomrule
    \end{tabular}}%
\end{table}

From Table \ref{table:Linear-Reg}, we observe that our proposed method achieves the best detection performance across all scenarios. As expected, the original NOT method is not well-suited for general linear regression systems. The NSP-Reg and NSP-Reg-Self methods perform similarly under normally distributed errors; however, consistent with observations in the linear system, NSP-Reg fails under $t$-distributed errors. Additionally, the reported distances $d(\mathcal{M}, \mathcal{T}^*)$ for our method remain within $6\%$ of the minimum spacing $\delta_n = 176$, further demonstrating its strong localization accuracy. Taken together, these simulation results confirm that the proposed method delivers robust FDR control, high TPR, and precise detection across a variety of settings.

\section{Real Data Analysis}
\label{sec:application}
\subsection{Spread of COVID-19 in Italy}
The novel coronavirus (COVID-19) is an emerging disease that has developed in the entire world and has attracted worldwide attention. We analyze the Italian COVID-19 dataset that is available from Johns Hopkins University Center for Systems Science and Engineering (JHU CCSE) Coronavirus (\url{https://systems.jhu.edu/research/public-health/ncov/}), consisting of daily summaries of the COVID-19 cases from 20th March 2020 to 8th August 2021. The dataset can be derived through the R package {\tt coronavirus}. The spread of COVID-19 can be modeled by the following well-known Susceptible-Infected-Recovered (SIR) epidemic equation \citep{SIR},
\begin{equation*}
        \frac{ds(t)}{dt} = -\beta i(t)s(t), \quad \frac{di(t)}{dt} = \beta i(t)s(t) - \gamma i(t),
\end{equation*}
where $s$ is the number of susceptible individuals, $i$ is the number of infectious individuals, $\beta$ is the infection rate, and $\gamma$ is the removal (recovery or death) rate. According to the government policies, medical resources, and some other external factors, the infection and removal rate are possibly different in different periods. Since the number of susceptible individuals can not be observed directly, we utilize the proposed method on the infection curve $i(t)$ for detecting locations of structural changes.

\begin{figure}[h]
\centering
\includegraphics[width = 15cm]{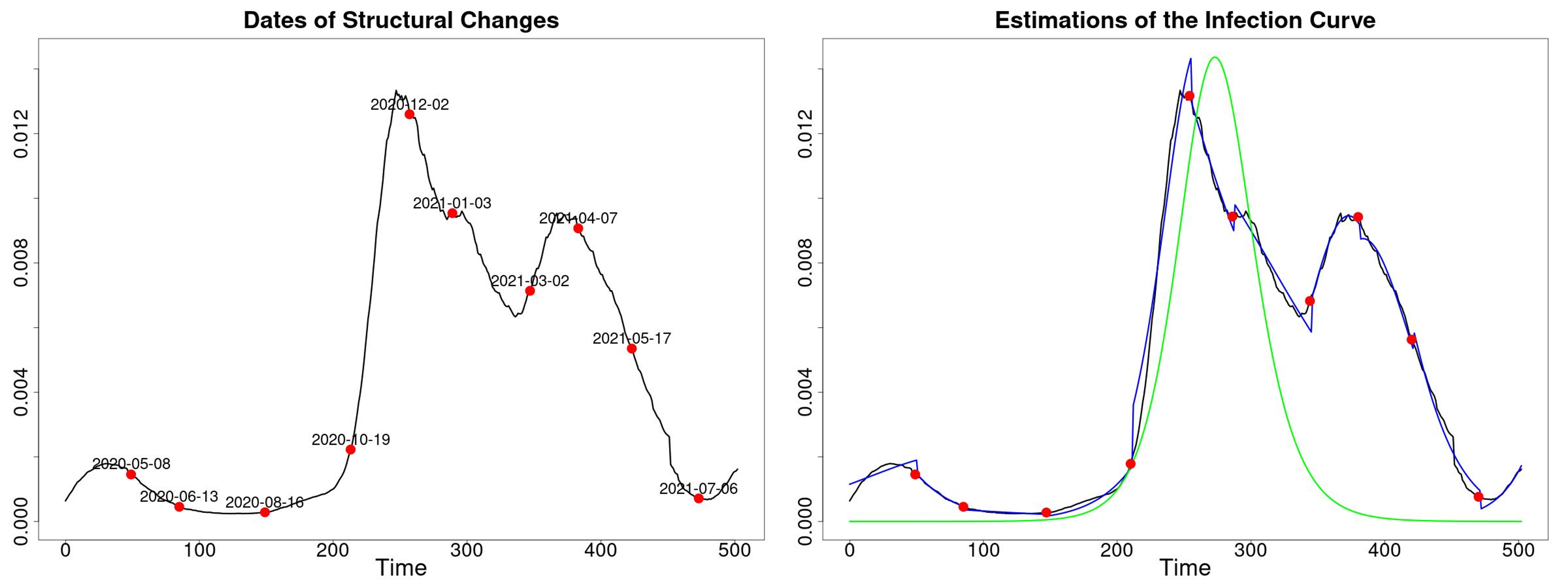}
\caption{\small Detection results and the number of infectious individuals of COVID-19 in Italy (black line). Left panel: the detected structural changes (red dot) and their corresponding date. Right panel: the estimate of the infection curve with zero change (green line) and that with detected structural changes (blue line).}
\label{fig:COVID19}
\end{figure}

The number of infectious individuals and the detected structural changes are shown in the left panel of Figure \ref{fig:COVID19}. We identify ten structural changes, which align closely with the major policy shifts and epidemic events in Italy. In particular, stringent control measures introduced in early 2020 reduced transmission by mid-April, and infections began to decline as medical resources improved. Restrictions were gradually relaxed during the summer, including the reopening of schools and domestic travel, and infections resurged in the autumn. After stricter measures were reinstated in mid-October, the outbreak was brought under control by early December. See \cite{Bosa22} for additional details on events and policies in 2020. With the rollout of vaccines at the beginning of 2021, infections declined again in the following months. Restrictions were then eased in early February 2021, contributing to another rebound from late February to late March, before renewed controls at the end of March mitigated transmission. Italy began reopening again in mid-2021, but infections increased during the summer with the spread of new variants.

Additionally, the estimates of $i(t)$ with zero change and detected structural changes are presented in the right panel of Figure \ref{fig:COVID19}. The fitted curves with zero change (green line) only capture one peak of the original data, while the curve with detected changes (blue line) fits the original data well, which also supports the detection results.

\subsection{Temperature change in Holocene}
We analyze the global temperature reconstruction for the Holocene (past 11,300 years) compiled by \cite{Marcott13}. The dataset provides a globally stacked temperature series constructed using their Standard method, which averages standardized proxy temperature records across sites after temporal alignment. To account for reconstruction uncertainty, we add independent Gaussian perturbations to the records based on their reported standard deviations. Both the reconstructed temperatures and their associated standard deviations are available in the supplementary data file “marcott.sm.database.s1.xlsx” accompanying \cite{Marcott13}, and can be accessed via \url{https://www.science.org/doi/suppl/10.1126/science.1228026/suppl_file/marcott.sm.database.s1.xlsx}.

The dynamic system we used here is the energy balance equation \citep{Kaper13},  
\begin{equation*}
R\frac{d T}{d t} = Q(1 - \alpha) - \varepsilon\sigma T^4,
\end{equation*}
where $T$ is the average temperature on the Earth's surface, $t$ is the time (year), $R$ is the average heat capacity of the Earth system, $Q$ is the annual global mean incoming solar radiation per square meter of the Earth's surface, $\alpha$ is the planetary reflectivity, $\varepsilon$ is the emission rate of the Earth and $\sigma = 5.67\times 10^{-8}$ is the Stefan-Boltzmann constant. To ensure identifiability, we set the average heat capacity $R = 20.83$ and the planetary reflectivity $\alpha = 0.3$ according to previous research \citep{Kaper13}.

\begin{figure}[h]
\centering
\includegraphics[width = 12cm]{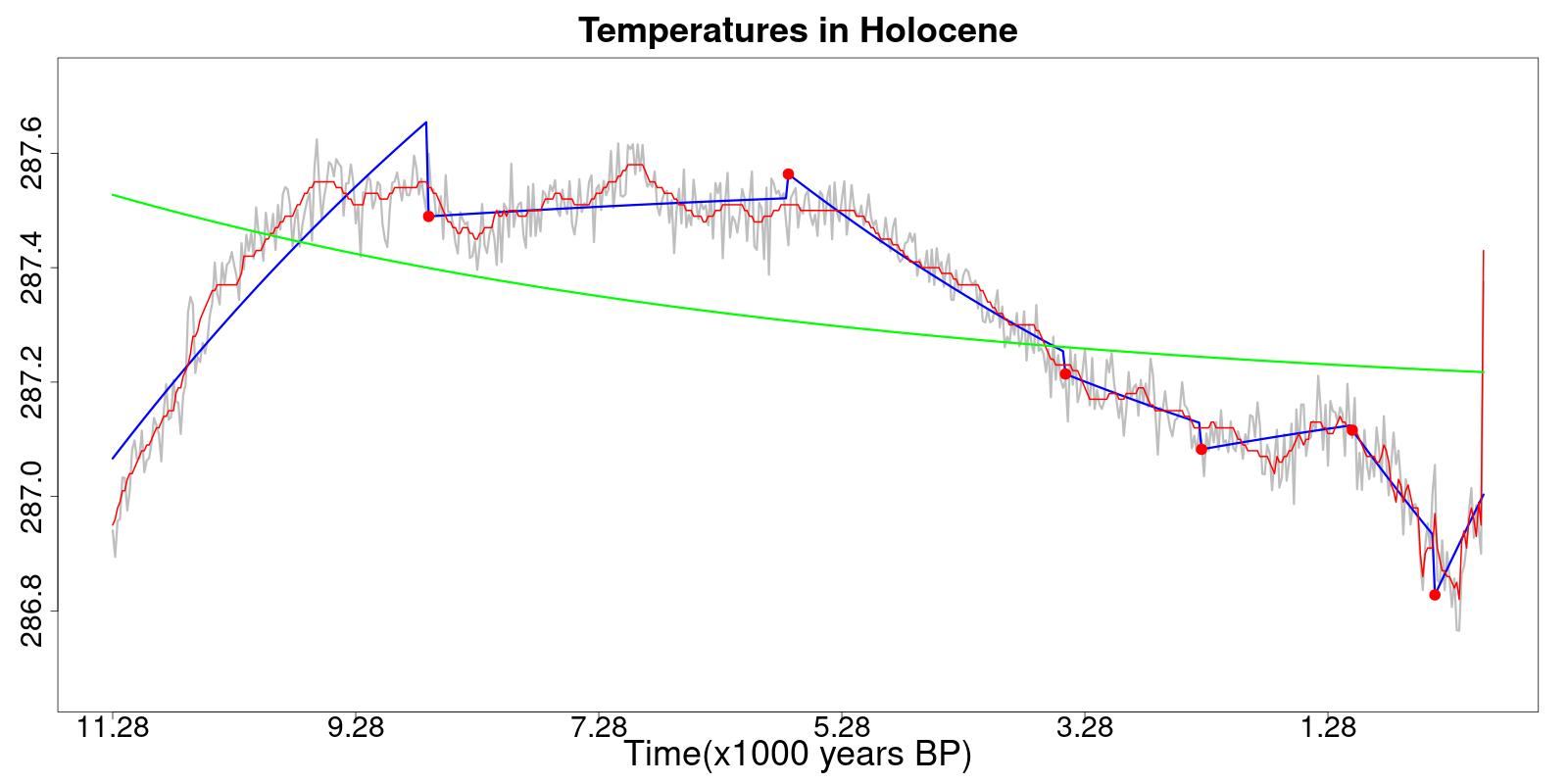}
\caption{\small Detection results and the temperatures in the Holocene. Red line: the original reconstruction of \cite{Marcott13}; grey line: the reconstruction that takes 1$\sigma$ uncertainty into account; red dot: detected structural changes; blue line: estimated temperatures with detected structural changes; green line: estimated temperatures with zero structural change.}
\label{fig:Temp}
\end{figure}

Figure \ref{fig:Temp} shows the temperatures in the Holocene, the detected six structural changes, and the corresponding estimates of the temperature curve. The six structural changes occurred around 8680, 5720, 3440, 2320, 1080, and 400 years before present (BP). These estimated periods align closely with major transitions identified in paleoclimate research. In particular, the early Holocene warming phase—known as the Holocene Climate Optimum (HCO, 9500–5500 BP; \citealp{Marcott13})—is clearly reflected in our results. The temperature increased gradually until around 8500 BP, remained relatively stable between 8500 and 5720 BP, and then entered a long-term cooling phase lasting roughly 3000 years. The subsequent warming trend beginning near 2000 BP corresponds to the Medieval Climate Optimum (MCO, 950–1250 BP; \citealp{Micheal09MCO}). A cooling transition detected around 1000 BP marks the onset of the coldest interval between 100 and 700 BP, consistent with the Little Ice Age (100–600 BP; \citealp{Matthews05}). Finally, the detected change around 400 BP indicates the beginning of a sustained warming trend, aligning with the onset of the Contemporary Climate period (\citealp{Thomas03MGCC}). Overall, the close correspondence between the detected structural changes and established geological periods supports the validity of our proposed method in identifying meaningful regime shifts in paleoclimate dynamics.

\vspace{-0.2in}
\section{Discussion}
\label{sec:discussion}
\vspace{-0.2in}
In this paper, we develop a unified framework for detecting structural changes in dynamic systems governed by ordinary differential equations. The method combines residual-based discrepancy and normalized parameter contrasts, uses the SNOT algorithm with a data-driven threshold for candidate screening, and applies an SPC procedure for FDR control based on order-preserving sample splitting. It accommodates both stable and diverging trajectories, applies to nonlinear dynamics without requiring closed-form ODE solutions, and is supported by theoretical guarantees on detection consistency, localization accuracy, and FDR control. Several directions remain for future work. These include extensions to broader classes of dynamic systems, such as stochastic, partial, and delay differential equations, as well as scalable adaptations to more general growth settings and high-dimensional models.

\vspace{-0.2in}
\section{Disclosure statement}\label{disclosure-statement}
\vspace{-0.2in}
No competing interest is declared.
\vspace{-0.2in}

\section{Data Availability Statement}\label{data-availability-statement}
\vspace{-0.2in}
\textcolor{black}{Previously published data were used for this work. The Italian COVID-19 dataset is available from Johns Hopkins University Center for Systems Science and Engineering (JHU CCSE) Coronavirus (\url{https://systems.jhu.edu/research/public-health/ncov/}). The global temperature reconstruction dataset for the Holocene can be accessed via \url{https://www.science.org/doi/suppl/10.1126/science.1228026/suppl_file/marcott.sm.database.s1.xlsx}.}
\vspace{-0.2in}



\vspace{0.2in}

\begin{center}

{\large\bf SUPPLEMENTARY MATERIAL}
\end{center}
\vspace{-0.2in}
\begin{description}
\item[Supplement] Algorithms, Theoretical details, along with additional simulation settings and results for ``Structural Change Detection in Dynamic Systems''. (.pdf file)
\end{description}
\vspace{-0.2in}

\small
\bibliography{bibliography.bib}

\end{document}